\pdfoutput=1

\documentclass[a4paper]{spie} 

\usepackage{amsmath,amsfonts,amssymb}
\usepackage{graphicx}
\usepackage[colorlinks=true, allcolors=blue]{hyperref}

\usepackage{fontawesome}

\usepackage{pdfcomment}
\usepackage[svgnames,table]{xcolor}
\usepackage{booktabs}
\definecolor{cy}{rgb}{1.00,0.831,0.474} 

\title{Simulating the radiation loss of superconducting submillimeter wave filters and transmission lines using Sonnet EM
}

\author[a,b,*]{Akira~Endo}
\author[a,c]{Alejandro~Pascual~Laguna}
\author[a,c]{Sebastian~H\"ahnle}
\author[a,c]{Kenichi~Karatsu}
\author[a]{David~J.~Thoen}
\author[c]{Vignesh~Murugesan}
\author[a,c]{Jochem~J.A.~Baselmans}
\affil[a]{Faculty of Electrical Engineering, Mathematics and Computer Science, Delft University of Technology, Mekelweg 4, 2628 CD Delft, the Netherlands.}
\affil[b]{Kavli Institute of NanoScience, Faculty of Applied Sciences, Delft University of Technology, Lorentzweg 1, 2628 CJ Delft, The Netherlands.}
\affil[c]{SRON---Netherlands Institute for Space Research, Sorbonnelaan 2, 3584 CA Utrecht, The Netherlands.}

\authorinfo{Further author information: A.E.: E-mail: a.endo@tudelft.nl}

\renewcommand{\textmu}{$\mathrm{\mu}$}

\begin{document} 
\maketitle

\begin{abstract}
Superconducting resonators and transmission lines are fundamental building blocks of integrated circuits for millimeter-submillimeter astronomy. 
Accurate simulation of radiation loss from the circuit is crucial for the design of these circuits because radiation loss increases with frequency, and can thereby deteriorate the system performance.
Here we show a stratification for a 2.5-dimensional method-of-moment simulator \textsf{Sonnet EM} that enables accurate simulations of the radiative resonant behavior of submillimeter-wave coplanar resonators and straight coplanar waveguides (CPWs). The \textsf{Sonnet} simulation agrees well with the measurement of the transmission through a coplanar resonant filter at 374.6~GHz. Our \textsf{Sonnet} stratification utilizes artificial lossy layers below the lossless substrate to absorb the radiation, and we use co-calibrated internal ports for de-embedding. With this type of stratification, \textsf{Sonnet} can be used to model superconducting millimeter-submillimeter wave circuits even when  radiation loss is a potential concern.
\end{abstract}


\section{INTRODUCTION}

 Superconducting microwave resonators are used in astronomical instrumentation\cite{2019NatAs.tmp..418E, Rantwijk2016, 2011ApJS..194...24M}, quantum computation\cite{10.1038/nature13171, doi:10.1063/1.4919761} and solid state physics\cite{PhysRevLett.109.107003}. For designing the resonators and the superconducting transmission line circuit around it, the planar method of moment (MoM) simulation software \textsf{Sonnet em}\cite{sonnet_user_guide} (Sonnet, hereafter) is widely used. This is because \textsf{Sonnet} requires less simulation time compared to full 3D simulators (e.g., \textsf{CST Microwave Studio}\cite{CSTMicrowaveStudio} and \textsf{HFSS}\cite{HFSS}) for planar structures\cite{sonnet_how_EM_works}, and it is straightforward to model superconductors in \textsf{Sonnet}\cite{10.1063/5.0005047, 2012ITMTT..60.1235N}. 

In recent years, there is an increasing demand in astronomical instrumentation for superconducting circuits that operate in the millimeter-submillimeter (mm-submm) band up to 1~THz\cite{2019NatAs.tmp..418E,Wheeler:2018cg}. At above $\sim$100~GHz, radiation losses become significant for coplanar structures, and the quality factor of resonators can deteriorate\cite{2019JATIS...5c5004E}. Whilst full 3D simulators can simulate free-space boundaries, \textsf{Sonnet} requires a perfect electrical conductor (PEC) surface on the four walls perpendicular to the planar structures. This can lead to problems such as suppression of radiative modes, reflections from the walls back to the circuit, and standing waves. Recently it has been reported that the radiation loss of a straight coplanar waveguide (CPW) can be accurately modeled with \textsf{Sonnet} by using a box that is sufficiently large to allow all radiative modes to be excited, and a stratification that prevents reflection from the walls of the simulation box\cite{10.1063/5.0005047}. Here we present the details of such a simulation, and show that the same stratification can also be applied to account for radiation loss in \textsf{Sonnet} simulations of submillimeter wave resonators. We do this by comparing \textsf{Sonnet} simulations with experiments, and a simulation using \textsf{CST Microwave Studio} (\textsf{CST}, hereafter). 

For the \textsf{Sonnet} simulations we have used \textsf{Sonnet} Version 17.56 on a Windows Server 2019 workstation with two AMD~EPYC~7302 16-core processors (3.0~GHz), with hyper-threading (64 threads in total), and 512~GB of random access memory (16$\times$36~GB). 


\section{Superconducting SUBMILLIMETER WAVE Resonator}

\begin{figure}[thb]
  \centering
  \includegraphics[width=85mm]{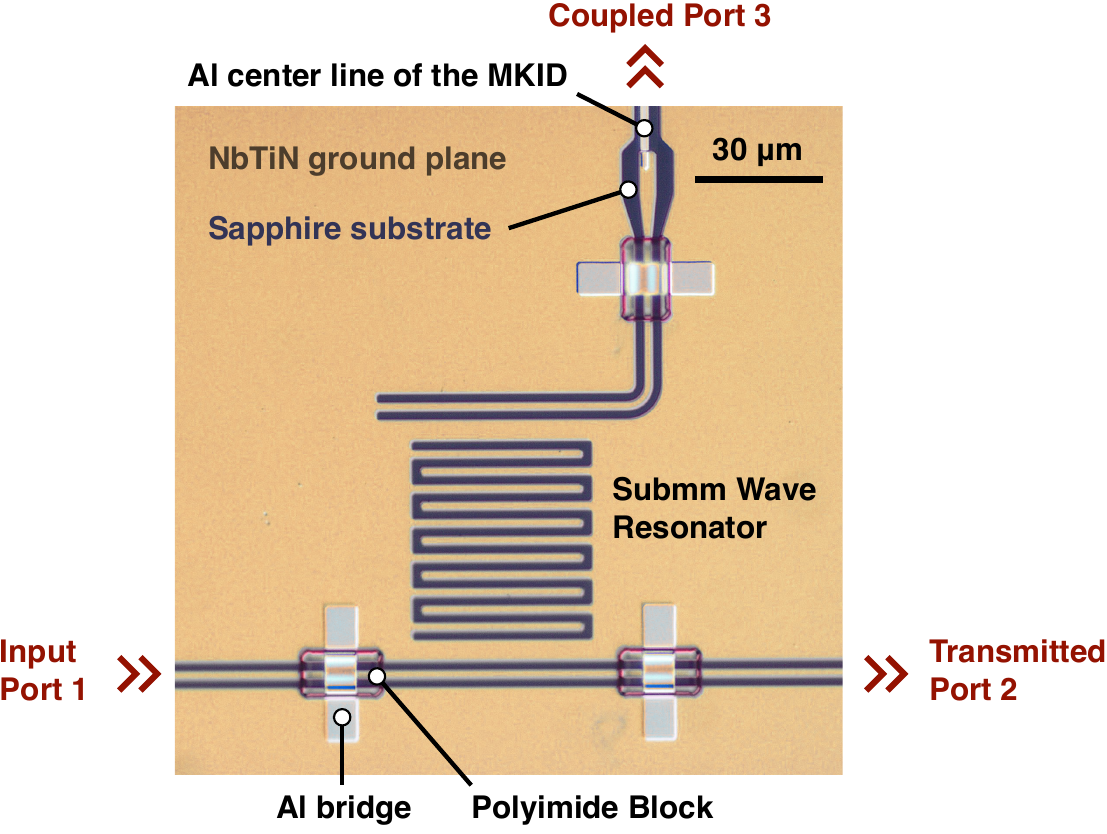}
  \caption{Micrograph of the 374.6~GHz superconducting resonator\cite{2019JATIS...5c5004E}.}
  \label{filter_micrograph}
\end{figure}

We will study the submm wave resonator shown in Fig.~\ref{filter_micrograph}. The resonator is a meandering slotline patterned in the superconducting NbTiN ground plane on a $c$-plane sapphire substrate. The second harmonic mode of this resonator at 374.6~GHz is used as one of the bandpass filters in a filter bank circuit of the astronomical spectrometer DESHIMA (Deep Spectroscopic High-redshift Mapper)\cite{2019JATIS...5c5004E}. At this frequency, the filter intercepts the submm wave signal flowing from the input \textsf{port 1} to the transmission \textsf{port 2}, so that part of the signal is directed to the coupled \textsf{port 3}. The power sent to the coupled \textsf{port 3} is measured with a microwave kinetic inductance detector (MKID). A previous study~\cite{2019JATIS...5c5004E} of this resonator has shown that, at the resonance frequency, $\sim$13\% of the power from the input is radiated into the substrate. This radiative behavior makes it challenging to reproduce these results using \textsf{Sonnet}.

\begin{figure}[thb]
  \centering
  \includegraphics[width=85mm]{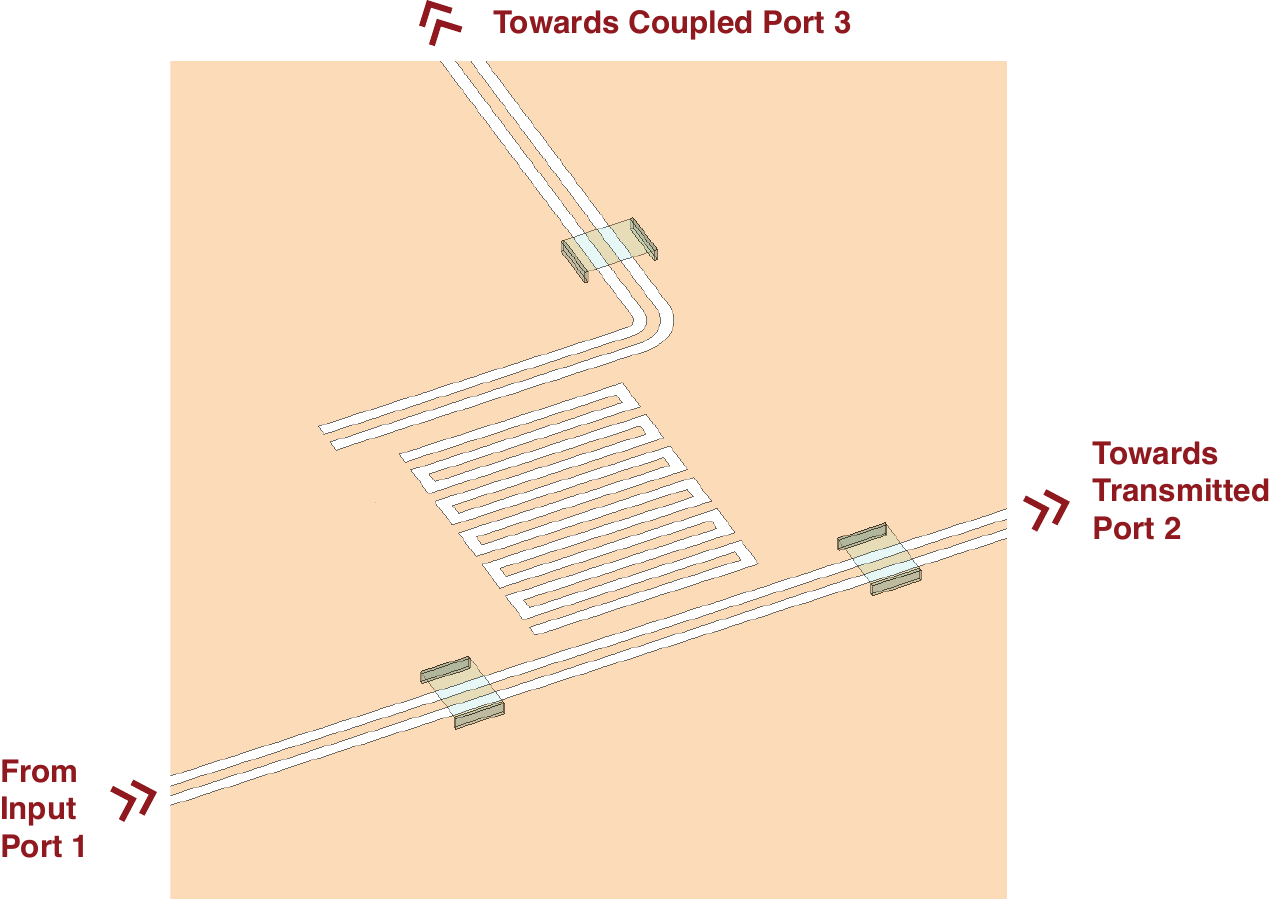}
  \caption{Design of the submm resonator in \textsf{Sonnet}. The three CPW lines extend to the edge of the ground plane, where they are terminated with an impedance-matched port.}
  \label{filter_sonnet_3d}
\end{figure}

In Fig.~\ref{filter_sonnet_3d} we show the model of the resonator in \textsf{Sonnet}. All slots in the NbTiN film are 2-\textmu m wide. The 100-nm thick NbTiN film is modeled as a metal sheet with a surface inductance of 1.0~pH/$\square$, calculated from the resistivity (102~$\mathrm{\mu}\Omega$~cm), critical temperature (14.7~K), and thickness (100~nm) of a NbTiN film deposited in an identical manner\cite{10.1063/5.0005047}. (Note that the calculation of the surface inductance requires the full Mattis-Bardeen equations, and not the low-frequency limit of $L_\mathrm{s}=R_\mathrm{s}\hbar/(\pi\Delta)$, where $R_\mathrm{s}$ is the sheet resistance, $\hbar$ is the reduced Planck constant, and $\Delta$ is the superconducting gap energy.) The polyimide blocks under the aluminium bridges have been replaced with a vacuum layer to save computation time. The aluminium film used for the bridges is modeled as a metal sheet with a surface resistance of $0.63\ \Omega/\square$. The CPW lines have a characteristic impedance of $93\ \Omega$. The CPW lines extend straight to the ports loaded with a matched impedance. The sapphire substrate is modeled with an anisotropic permittivity of 9.3 in the horizontal directions, and 11.9 in the vertical direction.

\section{Sonnet Stratification}

\subsection{Box}

\begin{figure}[thb]
  \centering
  \includegraphics[width=\textwidth]{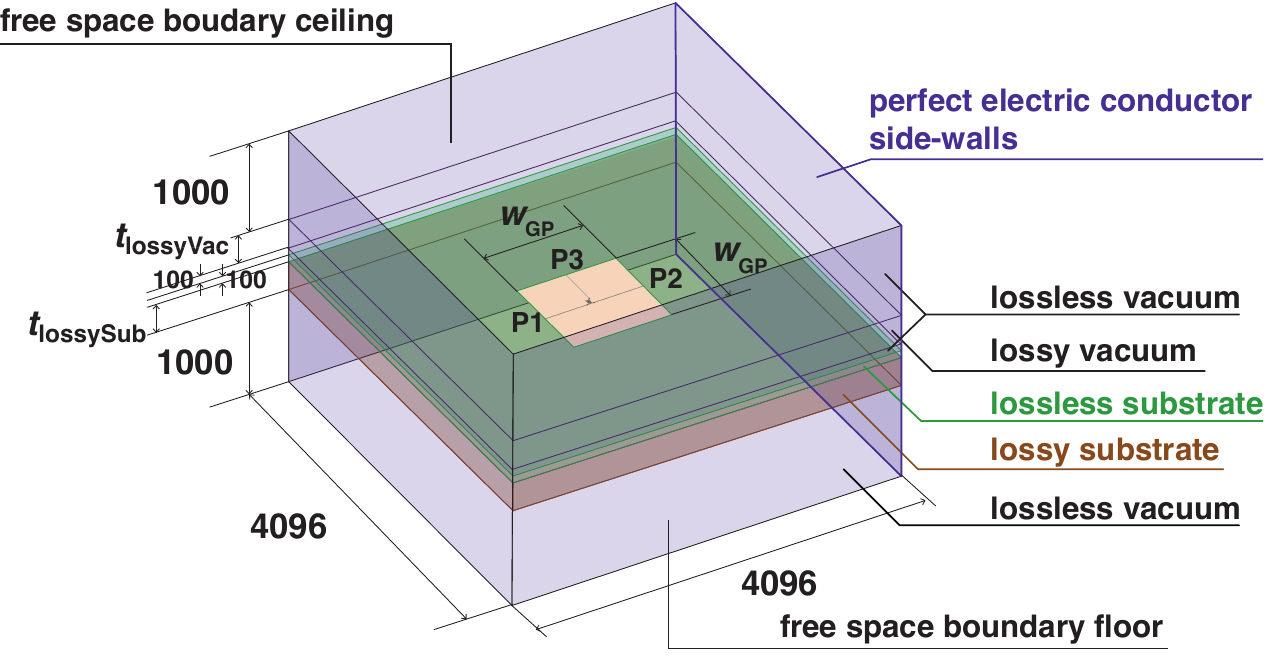}
  \caption{Stratification for the \textsf{Sonnet} simulation.}
  \label{sonnet_box}
\end{figure}

Fig.~\ref{sonnet_box} shows the stratification of the ground plane, dielectric layers and the box around the resonator. In Sonnet, the four sidewalls of the box must be made of PEC. The ceiling and floor are set to the `free space' boundary condition. In the horizontal directions, the box is 4096~\textmu m by 4096~\textmu m wide. The box is much larger than the metallization, for the following reasons:

\begin{enumerate}
	\item  When the sidewalls are close to each other, the box behaves like a rectangular waveguide. This means that radiation from the structure will be restricted to the set of discrete waveguide modes. This can lead to suppression of radiative modes that can be excited if the structure were placed in an infinite space. The typical scale recommended in the \textit{Sonnet User's Guide} is one or two times the wavelength\cite{sonnet_user_guide}.
	\item  When the sidewalls are not far enough from the structure, the radiation can be reflected at the walls and couple back into the structure. To prevent this, there must be sufficient attenuation of the signal between the structure and the walls. When standing waves are observed in the current distribution, this is a strong indication that there is significant reflection.
	\item  The area of the ground plane is kept to the minimum, because the calculation time of \textsf{Sonnet} scales roughly with the number of metal subsections cubed\cite{sonnet_how_EM_works}.
\end{enumerate}

In order to model an infinite space seen by the structure, we have introduced: 1) a double-layer substrate that is designed to absorb the radiated power without affecting the waves in the resonator and CPWs. Directly under the NbTiN film layer, there is a 100-\textmu m thick layer of sapphire with no dissipative loss. The thickness of 100~\textmu m is much larger than the slot width of 2~\textmu m, ensuring that all the guided waves (near field) are contained in this layer. Under this lossless substrate, there is an artificial lossy substrate which has the same anisotropic permittivity of $c$-plane sapphire and a loss tangent of $\tan \delta=0.1$. The small difference in complex permittivity between the lossless and lossy layers ensures that there is little reflection at the interface. The thickness of the lossy substrate, $t_\mathrm{lossy}$, will be chosen so that the signal radiated from the circuit is strongly attenuated and is not reflected back to the structure. For the same reason, we have also placed an artificial layer of lossy vacuum at 100~\textmu m above the ground plane layer to absorb radiation that is launched above the substrate, depending on the circuit that is modeled. As we show in subsection~\ref{subsec:lossyvac}, this layer has no effect for the cases simulated in this paper. Finally, above the lossy vacuum layer and below the lossy substrate, there are 1-mm thick layers of lossless vacuum that separate the lossy layers from the free-space boundaries at the top and bottom of the box.

\subsection{Ports}

Here we use `co-calibrated internal ports', because this type of port can be placed on the edge of a metal that is not attached to the \textsf{Sonnet} box. (The most commonly used `box-wall port' must be attached to the box wall.) The co-calibrated internal ports are de-embedded, so the results are accurate\cite{sonnet_user_guide}. For our simulations we apply a `floating' ground node connection, which internally creates a `Generalized Local Ground' (GLG) metal connection at every port\cite{sonnet_user_guide}. This GLG metal is removed in the de-embedding process. The short calibration length of 6~\textmu m follows from a private discussion with Sonnet, but it is not studied in detail by us.

\section{Results}

\subsection{`Reference' Geometry}

\rowcolors{2}{White}{LightBlue!30}
\begin{table}[htbp]
\centering
\begin{tabular}{r|l}\toprule
$t_\mathrm{lossySub}$ & $10^2$~\textmu m \\
$t_\mathrm{lossyVac}$ & $10^2$~\textmu m \\
$\tan \delta$ & $10^{-1}$ \\
$w_\mathrm{GP}$ & 1024~\textmu m \\
$\Delta_x$ & 0 \\
$\Delta_y$ & 0 \\
box size & 4096~\textmu m $\times$ 4096~\textmu m \\\bottomrule
\end{tabular}
\caption{Parameters defining the \textit{reference} geometry. $t_\mathrm{lossySub}$ and $t_\mathrm{lossyVac}$ are the thicknesses of the artificial lossy layers in vacuum and below the substrate, respectively (see Fig.~\ref{sonnet_box}). 
$\tan \delta$ is the loss tangent of those layers. $w_\mathrm{GP}$ is the width of the ground plane (see Fig.~\ref{sonnet_box}). 
$\Delta_x$ and $\Delta_y$ are the offsets of the resonator position, see Fig.~\ref{pic:xypos}. Box size is the lateral dimension of the box (see Fig.~\ref{sonnet_box}).
}
\label{tab:ref}
\end{table}

\begin{figure}[thb]
  \centering
  \includegraphics[width=130mm]{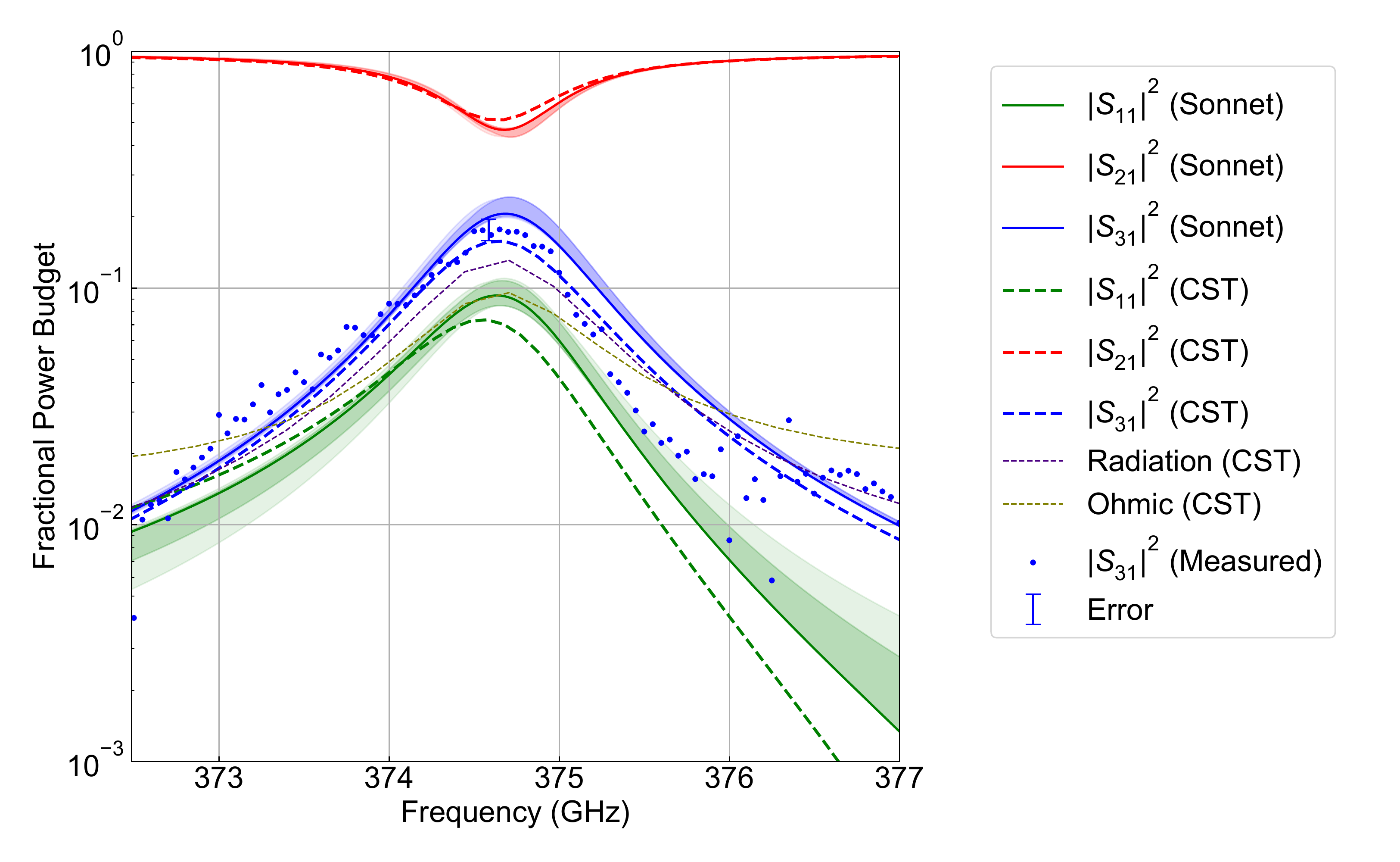}
  \caption{Fractional power budget as a function of frequency calculated by \textsf{Sonnet} (solid curves) using the \textit{reference} geometry, compared to the result of \textsf{CST} (dashed curves) and measurement\cite{2019JATIS...5c5004E} (points). Shaded regions which indicate the maximum-minimum range of $|S_{n1}|^2$ curves calculated with position offsets indicated in Fig.~\ref{pic:xypos} and Table~\ref{tab:xypos}. The thicker shading indicates the central 3$\times$3 locations, and the thinner shading indicates all 5$\times$5 locations.}
  \label{plot_xypos}
\end{figure}

\begin{figure}[thb]
  \centering
  \includegraphics[width=65mm]{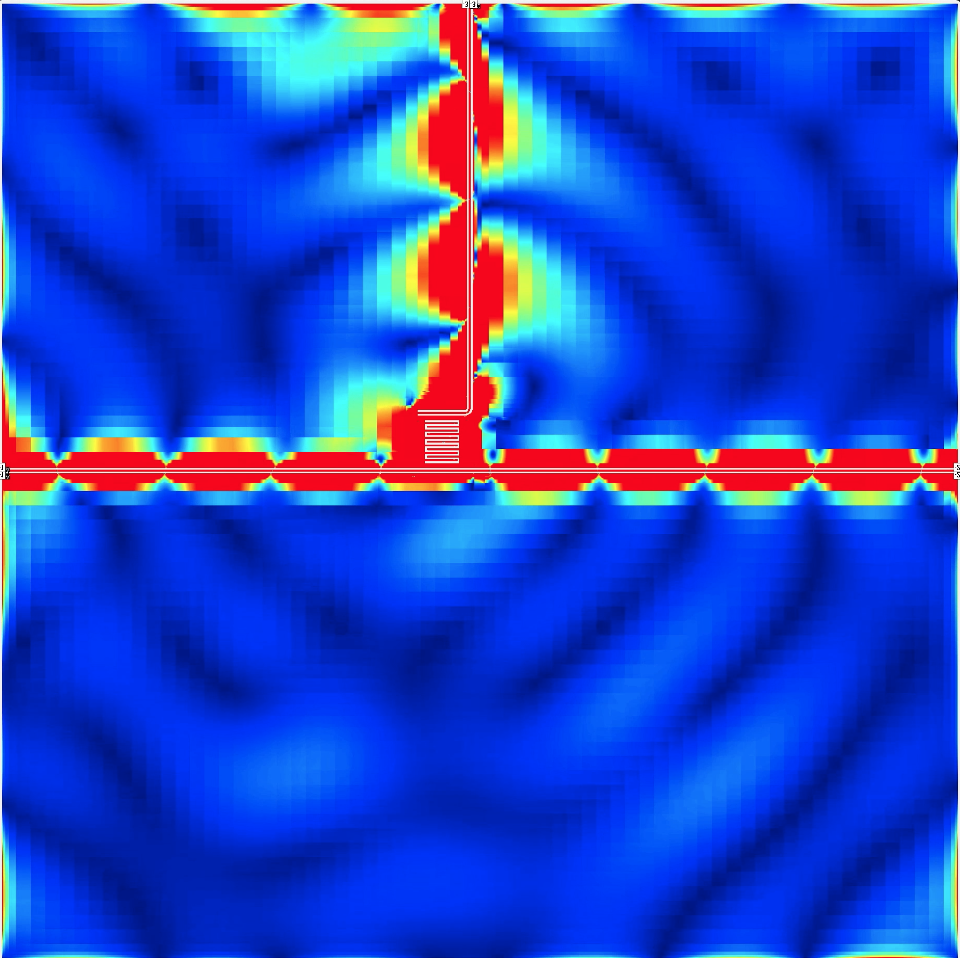}
  \caption{Snapshot of the current distribution $J_{xy}$ for the \textit{reference} geometry (\href{https://youtu.be/Vu4y0WdD9tI}{\faYoutubePlay \ \textsf{online video}}). Note that the current densities plotted do not represent de-embedded data, and therefore areas near any port include the effect of the port discontinuity, according to the manual of \textsf{Sonnet}\cite{plotting_conventions}.}
  \label{ref_current}
\end{figure}

\begin{figure}[thb]
  \centering
  \includegraphics[width=65mm]{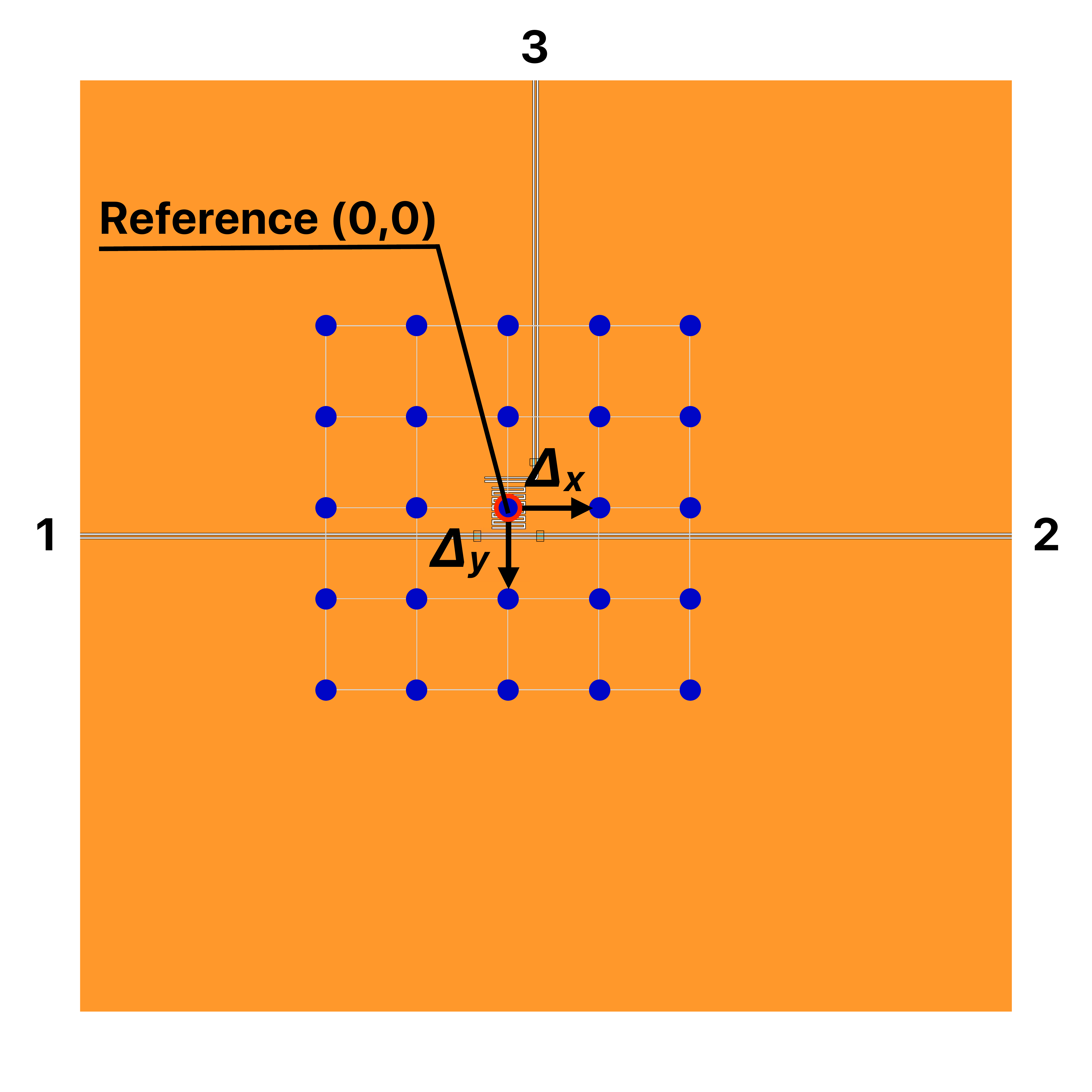}
  \caption{Locations of the \textit{reference} and offset positions on the 1024~\textmu m $\times$ 1024~\textmu m ground plane. The numbers 1--3 indicate the location of the ports.}
  \label{pic:xypos}
\end{figure}

We will first discuss the \textit{reference} geometry, of which the parameters are listed in Table~\ref{tab:ref}. The \textit{reference} geometry is chosen as a good compromise between accuracy and calculation time. The resonator is placed close to the center of the ground plane, which is 1024~\textmu m wide in both $x$ and $y$ directions. As mentioned earlier, for the NbTiN film in \textsf{Sonnet} we took a sheet inductance of 1.0~pH$/\square$ that is derived from measured film properties. However, we adjust the sheet inductance in \textsf{CST} to 0.73~pH$/\square$ to get the resonance frequency close to the measurement. The exact reason for the different sheet inductance required in \textsf{CST} is not clear at the time of publication.
Furthermore, in order to align the simulated resonance frequency to the measurement, we multiplied the \textsf{Sonnet} frequency by a factor of $1.0067$, and the \textsf{CST} frequency by a factor of $0.9986$.~\footnote{This was much easier than iterating the simulations with varying sheet inductance to reach virtually the same result.} 

The $S$-parameters calculated with \textsf{Sonnet} are compared with \textsf{CST} and measurements in Fig.~\ref{plot_xypos}. Because  we can obtain only $|S_{31}|^2$ from the measurements, we take $|S_{31}|^2$ as the prime indicator for the accuracy of the simulations. Both $|S_{31}|^2$ curves of \textsf{Sonnet} and \textsf{CST} are within the 1$\sigma$ error range of the measurement\cite{2019JATIS...5c5004E} at the resonance peak.

The time-evolution of the current is a good indication of the presence of standing waves in the simulation box. Fig.~\ref{ref_current} is a snapshot of an animation (\href{https://youtu.be/Vu4y0WdD9tI}{\faYoutubePlay \ \textsf{online video}}). The radiation leaving the resonator is not returning to the CPW, and there are no obvious indications of standing waves. (Typically, standing waves are visible if the box is too small, there is not enough dissipation in the box, etc.)

\subsection{Sensitivity to offsets of the resonator within the ground plane}

\rowcolors{2}{White}{LightBlue!30}
\begin{table}[htbp]
\centering
\begin{tabular}{r|l}\toprule
$t_\mathrm{lossySub}$ & $10^2$~\textmu m \\
$t_\mathrm{lossyVac}$ & $10^2$~\textmu m \\
$\tan \delta$ & $10^{-1}$ \\
$w_\mathrm{GP}$ & 1024~\textmu m \\
$\Delta_x$ & $-$200, $-$100, \textbf{0}, +100, +200~\textmu m \\
$\Delta_y$ & $-$200, $-$100, \textbf{0}, +100, +200~\textmu m \\
box size & 4096~\textmu m $\times$ 4096~\textmu m \\\bottomrule
\end{tabular}
\caption{Parameters defining the set of geometries used to test the sensitivity to offsets of the resonator within the ground plane. Bold characters indicate the \textit{reference} geometry.}
\label{tab:xypos}
\end{table}

In Fig.~\ref{plot_xypos} we show the effect of shifting the position of the resonator within the ground plane as indicated in Fig.~\ref{pic:xypos} and Table~\ref{tab:xypos}. When we take the peak value of $|S_{31}|^2$, the ratio between maximum and minimum of all positions is 1.24, giving an indication of the accuracy of the \textsf{Sonnet} simulation. 

Interestingly, the \textit{reference} position located near the center of the ground plane (hence also the center of the simulation box) yields a peak $|S_{31}|^2$ that is the lowest among all positions. This could be an indication that there remains an effect of a finite distance to the box wall or the ground plane edge. Further investigation in the directions of:
\begin{enumerate}
	\item moving the ground plane together with the resonator inside the box and comparing the effect, and
	\item repeating the simulation with larger box sizes and larger ground plane sizes
\end{enumerate}
are recommended if one would want to further improve the simulation accuracy.

\subsection{Sensitivity to the ground plane width}

\rowcolors{2}{White}{LightBlue!30}
\begin{table}[htbp]
\centering
\begin{tabular}{r|l}\toprule
$t_\mathrm{lossySub}$ & $10^2$~\textmu m \\
$t_\mathrm{lossyVac}$ & $10^2$~\textmu m \\
$\tan \delta$ & $10^{-1}$ \\
$w_\mathrm{GP}$ & 512, \textbf{1024}, 1536, 2048~\textmu m \\
$\Delta_x$ & 0 \\
$\Delta_y$ & 0 \\
box size & 4096~\textmu m $\times$ 4096~\textmu m \\\bottomrule
\end{tabular}
\caption{Parameters defining the set of geometries used to test the sensitivity to the ground plane width. Bold characters indicate the \textit{reference} geometry.}
\label{tab:GPsize}
\end{table}

\begin{figure}[thb]
  \centering
  \includegraphics[width=130mm]{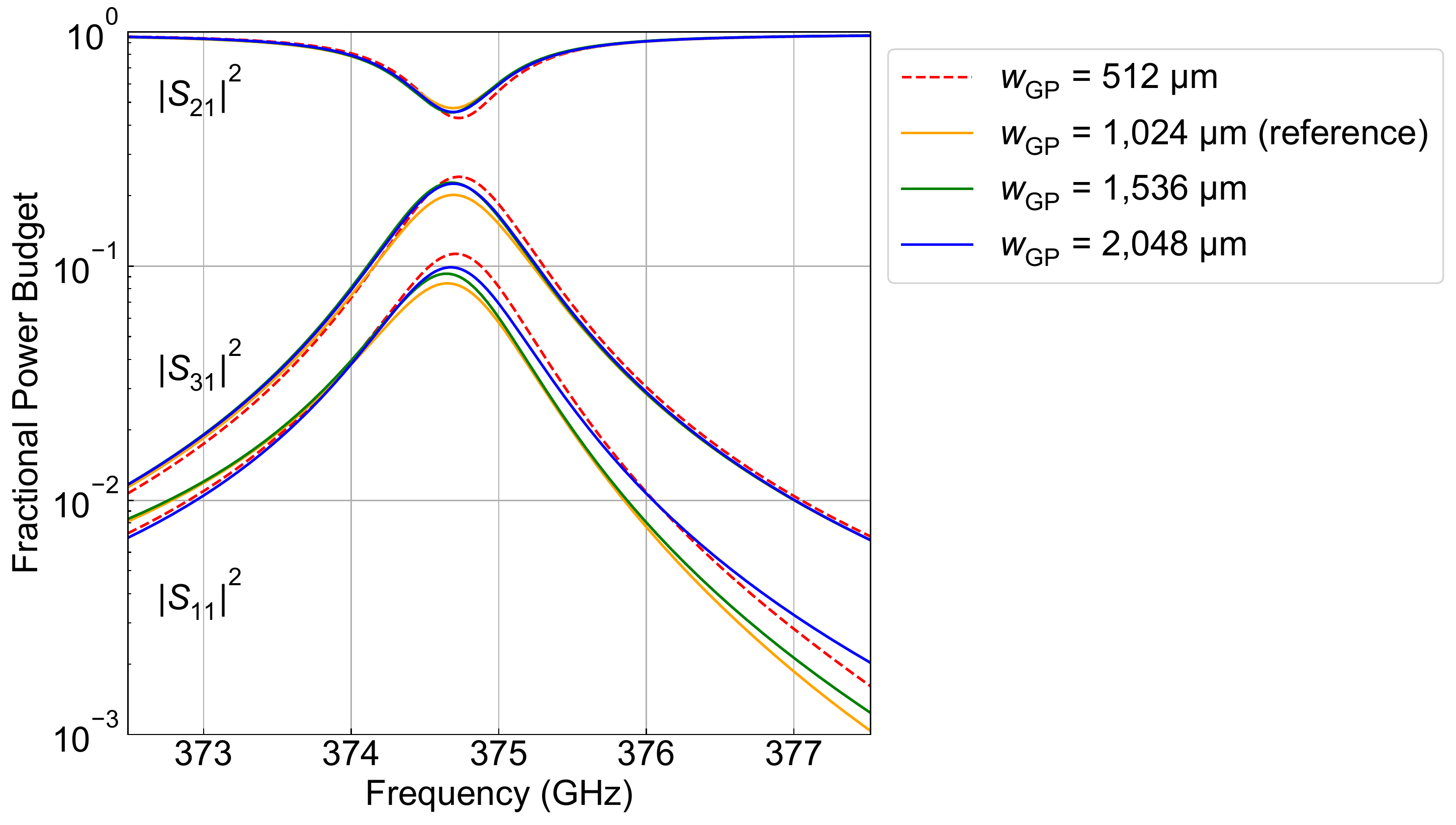}
  \caption{The effect of changing the ground plane width, $w_\mathrm{GP}$.}
  \label{plot:GPsize}
\end{figure}

The size of the ground plane can, in general, affect the amount of radiation loss. Here we keep the square shape of the ground plane and vary the length of the sides $w_\mathrm{GP}$ from 512~\textmu m to 2048~\textmu m, as summarized in Table~\ref{tab:GPsize}. The result is shown in Fig.~\ref{plot:GPsize}. Compared to the \textit{reference} geometry ($w_\mathrm{GP}=1024$~\textmu m), all other values for  $w_\mathrm{GP}$ yield a slightly higher peak $|S_\mathrm{31}|^2$ value. In general we expect that the larger the ground plane the more accurate the results are, because we are trying to simulate an infinite ground plane case. However, it should be noted that the experiment was also not with an infinite ground plane; there were neighboring filter channels on both sides of the filter at a distance of $3\lambda /4$. The $|S_\mathrm{31}|^2$ peak of the smallest and hence least accurate $w_\mathrm{GP}=512$~\textmu m geometry has a $|S_\mathrm{31}|^2$ peak that is 22\% higher than the \textit{reference}. The $|S_\mathrm{31}|^2$ peak of the largest, $w_\mathrm{GP}=2048$~\textmu m geometry has a $|S_\mathrm{31}|^2$ peak that is 13\% higher than the \textit{reference}. 

\subsection{Sensitivity to the thickness of the lossy substrate}

\rowcolors{2}{White}{LightBlue!30}
\begin{table}[htbp]
\centering
\begin{tabular}{r|l}\toprule
$t_\mathrm{lossySub}$ & $0$, $10^0$, $10^1$, $\mathbf{10^2}$, $10^3$, $10^4$~\textmu m \\
$t_\mathrm{lossyVac}$ & $10^2$~\textmu m \\
$\tan \delta$ & $10^{-1}$ \\
$w_\mathrm{GP}$ & 1024\textmu m \\
$\Delta_x$ & 0 \\
$\Delta_y$ & 0 \\
box size & 4096~\textmu m $\times$ 4096~\textmu m \\\bottomrule
\end{tabular}
\caption{Parameters defining the set of geometries used to test the sensitivity to the thickness of the lossy substrate. Bold characters indicate the \textit{reference} geometry.}
\label{tab:t_lossySub}
\end{table}

\begin{figure}[thb]
  \centering
  \includegraphics[width=130mm]{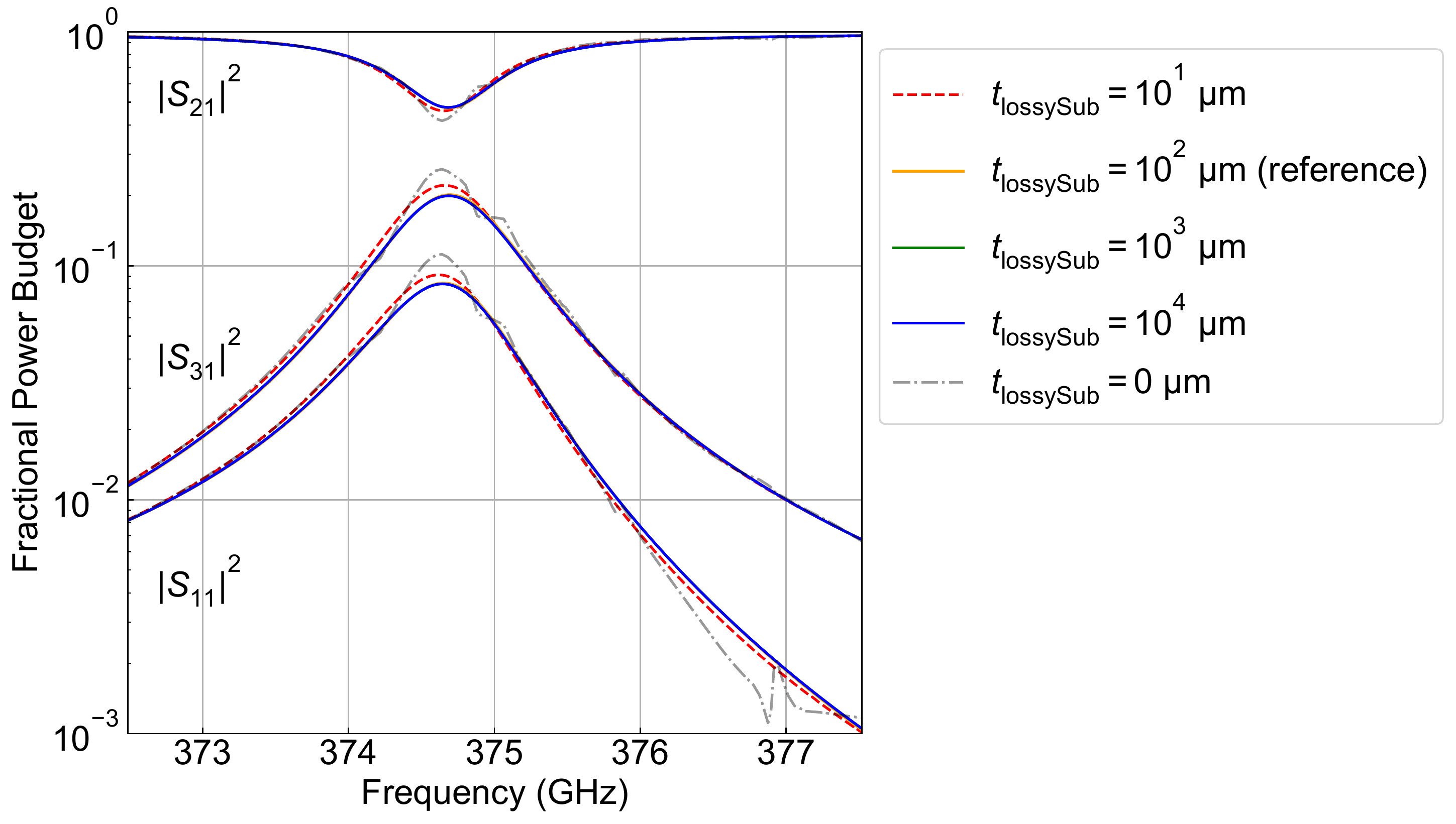}
  \caption{Effect of changing the thickness of the lossy substrate, $t_\mathrm{lossySub}$. The three curves for $t_\mathrm{lossySub}$ = $10^2$~\textmu m, $10^3$~\textmu m, and $10^4$~\textmu m are completely overlapping for all three $S$-parameters. The curve for $t_\mathrm{lossySub}$ = 0~\textmu m shows substrate-resonance features, especially at around 375~GHz and 377~GHz.}
  \label{plot:t_lossySub}
\end{figure}

The thickness of the artificially introduced lossy substrate, $t_\mathrm{lossySub}$, has been varied from 0 to $10^4$~\textmu m, as summarized in Table~\ref{tab:t_lossySub}. For the 0~\textmu m and 1~\textmu m cases, the \textit{Adaptive Band Synthesis} (ABS) of \textsf{Sonnet} had difficulties in converging after simulating at many frequencies, so eventually the simulation had to be terminated. This is most likely because there was a strong standing wave excited in the box because there was not enough attenuation under the ground plane. This shows the necessity to include this lossy layer, and we will focus on the results of $t_\mathrm{lossy}=10^1$--$10^4$~\textmu m. Fig.~\ref{plot:t_lossySub} shows the resonance peak for the different lossy-substrate thicknesses. Except for the case where the lossy substrate is the thinnest, $t_\mathrm{lossySub} = 10^1$~\textmu m, the results are completely insensitive to varying $t_\mathrm{lossySub}$. This is an additional, strong indication that the waves radiated from the vicinity of the resonator are strongly attenuated by the lossy layer, thereby not returning to the circuit and not creating standing waves that affect the results.

Although large values of $t_\mathrm{lossySub}$ hardly affect the simulation time, we have chosen a minimal value of $t_\mathrm{lossySub} = 10^2$~\textmu m as the \textit{reference} value because it is sufficient.

\subsection{Sensitivity to the thickness of the lossy vacuum layer}
\label{subsec:lossyvac}

\rowcolors{2}{White}{LightBlue!30}
\begin{table}[htbp]
\centering
\begin{tabular}{r|l}\toprule
$t_\mathrm{lossySub}$ & $10^2$~\textmu m \\
$t_\mathrm{lossyVac}$ & $0$, $10^1$, $\mathbf{10^2}$, $10^3$, $10^4$~\textmu m \\
$\tan \delta$ & $10^{-1}$ \\
$w_\mathrm{GP}$ & 1024\textmu m \\
$\Delta_x$ & 0 \\
$\Delta_y$ & 0 \\
box size & 4096~\textmu m $\times$ 4096~\textmu m \\\bottomrule
\end{tabular}
\caption{Parameters defining the set of geometries used to test the sensitivity to the thickness of the lossy vacuum layer. Bold characters indicate the \textit{reference} geometry.}
\label{tab:t_lossyVac}
\end{table}

\begin{figure}[thb]
  \centering
  \includegraphics[width=130mm]{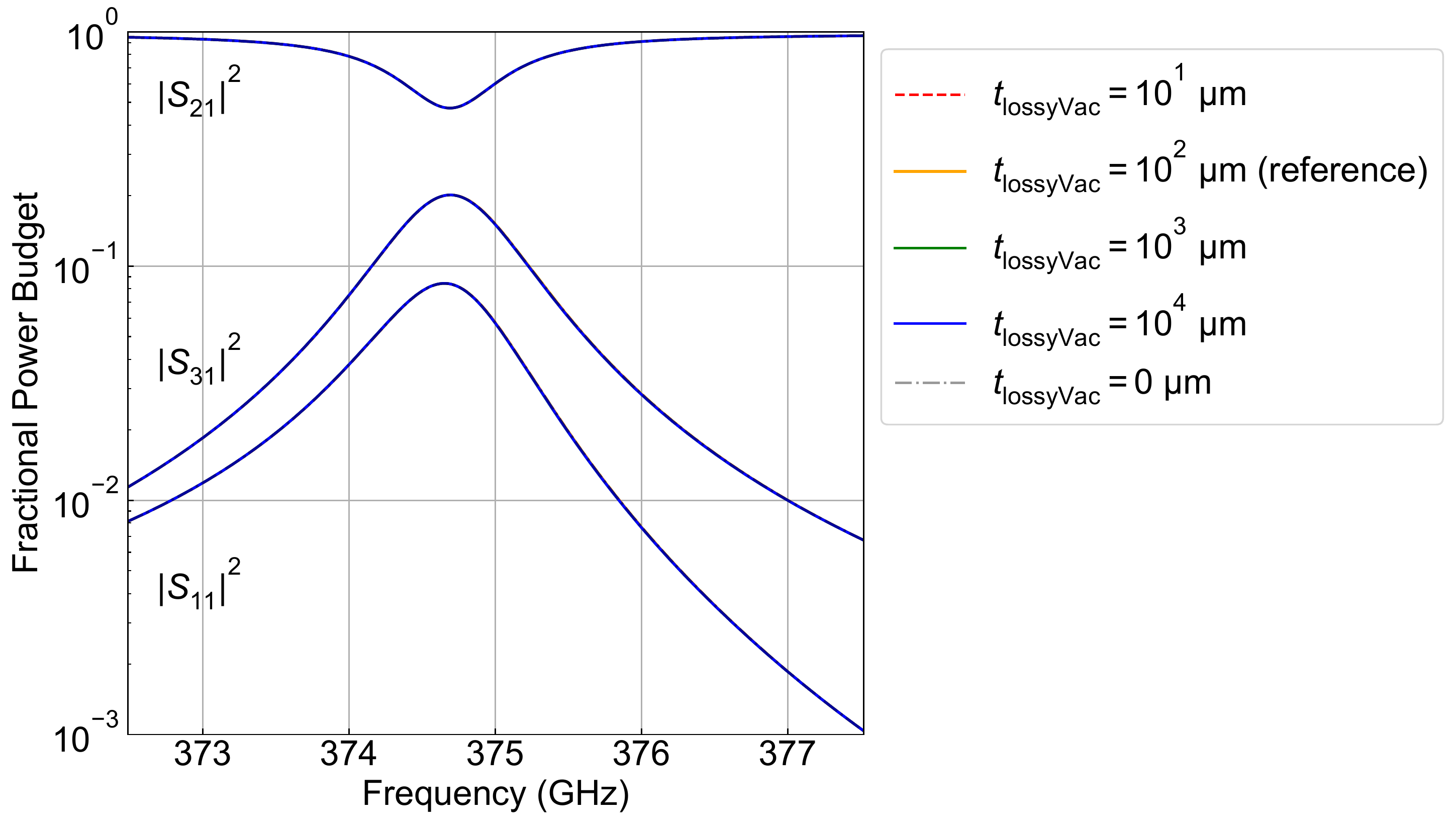}
  \caption{Effect of changing the thickness of the lossy vacuum layer, $t_\mathrm{lossyVac}$. The three curves for $t_\mathrm{lossyVac}$ = $10^2$~\textmu m, $10^3$~\textmu m, and $10^4$~\textmu m are completely overlapping for all three $S$-parameters.}
  \label{plot:t_lossyVac}
\end{figure}

In a similar way as the previous section, the thickness of the artificially introduced lossy vacuum layer $t_\mathrm{lossyVac}$ has been varied from 0 to $10^4$~\textmu m, as summarized in Table~\ref{tab:t_lossyVac}. The lossy vacuum layer has no noticeable effect on the results in the case of this model, as shown in Fig.~\ref{plot:t_lossyVac}. The lossy vacuum layer has no effect on the straight-CPW simulation in section~\ref{sec:straightCPW} either. This is as expected, because the radiation from these circuits is directed towards the substrate\cite{button1983infrared}. Moreover, even for geometries that might radiate towards the box ceiling, that radiation could be efficiently absorbed by the free-space boundary because there is no reflective dielectric-vacuum interface in between. Nevertheless, we include the lossy vacuum layer in the box to confirm that there are no standing waves in the volume above the substrate, by an analysis as shown in Fig.~\ref{plot:t_lossyVac}.

\subsection{Sensitivity to the cell size (meshing)}

\begin{figure}[thb]
  \centering
  \includegraphics[width=100mm]{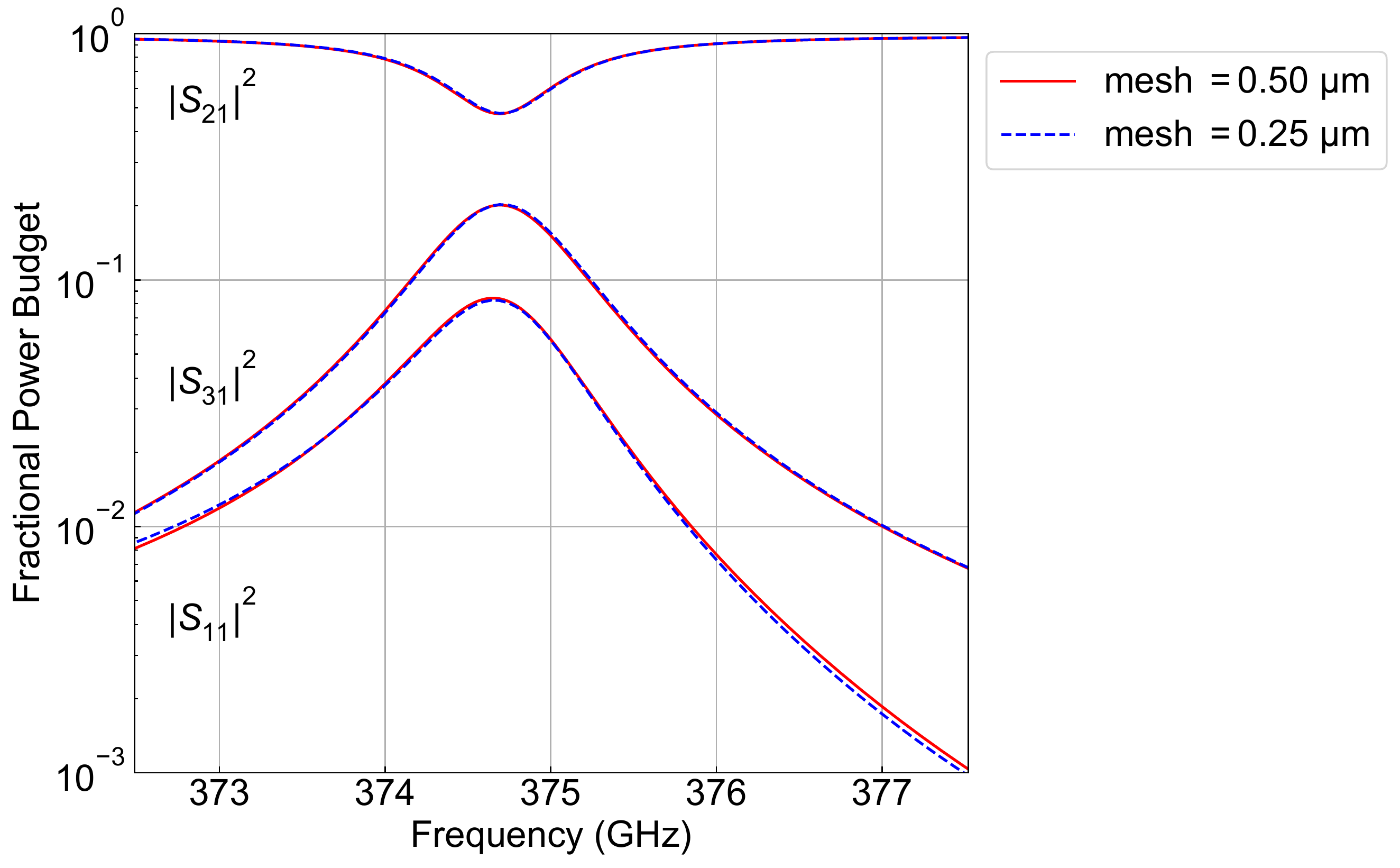}
  \caption{Effect of changing the mesh size, from the \textit{reference} value of 0.5~\textmu m (solid curves) to 0.25~\textmu m (dashed curves) in both $x$ and $y$ directions.}
  \label{plot:highres}
\end{figure}

Finally we check that the simulation has a sufficiently small cell size (or mesh size).
The \textit{reference} geometry had a cell size of 0.5~\textmu m in both $x$ and $y$ directions.
In Fig.~\ref{plot:highres} we compare the \textit{reference} result with a simulation in which the cell size was reduced to 0.25~\textmu m. Making the cell size smaller shifted the resonance frequency downwards by a factor of 1.0096. 
In Fig.~\ref{plot:highres} we have multiplied the frequency of the 0.25~\textmu m result with this factor so that the resonance frequencies match. After this correction, the two sets of curves are nearly overlapping. This result verifies that the \textit{reference} geometry is sufficiently meshed.

\section{Simulation of Radiation Loss from a Straight CPW}
\label{sec:straightCPW}

To show the general applicability of the stratification, we present the simulation of radiation loss from a straight CPW using the same stratification as the one used for the resonator. A similar simulation has been presented in Ref.\cite{10.1063/5.0005047}. Here we we will compare the results to the analytical formula of Ref.\cite{81658}. 

\begin{figure}[thb]
  \centering
  \includegraphics[width=100mm]{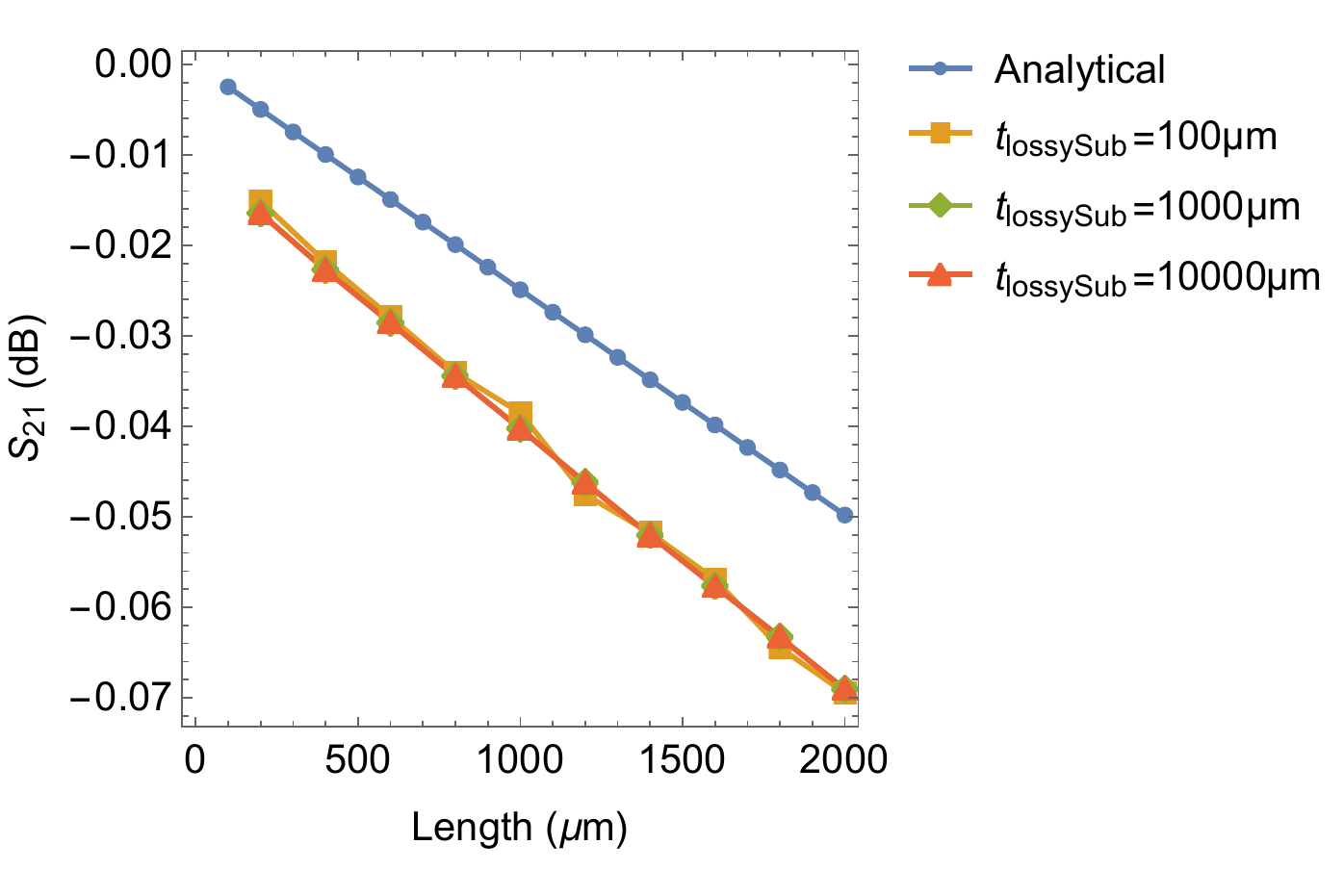}
  \caption{Radiative loss along the $s=w=2\ \mathrm{\mu m}$  CPW as a function of length. The analytical line is calculated using Eq. \ref{eq:Frankel}. The other curves are calculated using \textsf{Sonnet}, for different thicknesses of the lossy dielectric, $t_\mathrm{lossySub}$.
  }
  \label{plot:straightCPW}
\end{figure}

\begin{figure}[thb]
  \centering
  \includegraphics[width=100mm]{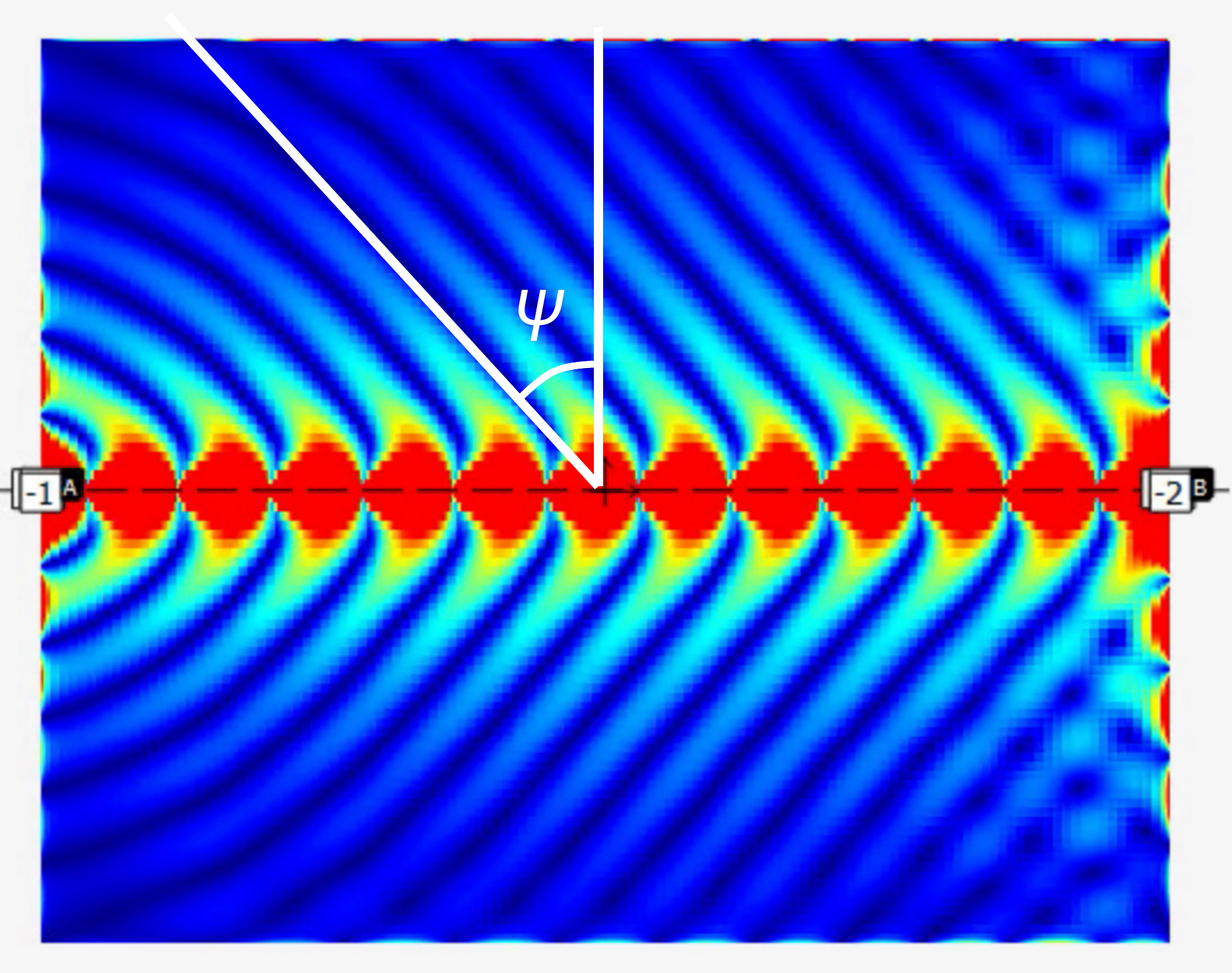}
  \caption{Snapshot of the current distribution $J_{xy}$ around a straight CPW (\href{https://youtu.be/iB3sx6GNrms}{\faYoutubePlay \ \textsf{online video}}). 
  $\Psi =  \arccos(\sqrt{\varepsilon_\mathrm{eff}/\varepsilon_\mathrm{r}}) = 42.5^\circ$ is derived from $\varepsilon_\mathrm{eff} = 6.22$ of the CPW line calculated by \textsf{Sonnet} and $\varepsilon_\mathrm{r}=11.44$ of the Si substrate. 
  Note that the current densities plotted do not represent de-embedded data, and therefore areas near any port include the effect of the port discontinuity, according to the manual of \textsf{Sonnet}\cite{plotting_conventions}. 
  }
  \label{cpw_animation}
\end{figure}

We have used \textsf{Sonnet} to simulate the radiation loss of a straight CPW made of PEC with a center strip width of 2~\textmu m and a slot width of 2~\textmu m, on a silicon substrate with a relative dielectric constant of 11.44. The lateral dimensions of the \textsf{Sonnet} box were 3000~\textmu m $\times$ 3000 \textmu m. The thicknesses of the layers were the same as those of the \textit{reference} geometry, except for the thickness of the lossy substrate that was varied: $t_\textrm{lossuSub}=10^2,\ 10^3,\ 10^4$ \textmu m. Fig. \ref{plot:straightCPW} shows the \textsf{Sonnet}-simulated $|S_{21}^2|$ for different CPW lengths\cite{10.1063/5.0005047}. $|S_{11}^2|$ was $<-45\ \mathrm{dB}$ with no clear frequency-dependence, which proves that the ports will matched well to the CPW line and reflections are negligible for this analysis. From the slope of a linear-fit to the $|S_{21}^2|$ as a function of frequency, we obtain a loss of 0.030 dB~$\mathrm{mm^{-1}}$. This is close to the loss of 0.025 dB~$\mathrm{mm^{-1}}$ obtained from analytical formula of Frankel et al.\cite{81658}:

\begin{equation}
\label{eq:Frankel}
\alpha_\mathrm{rad}=\bigl(\frac{\pi}{2}\bigr)^5 2 \Biggl( \frac{\sin^4\Psi}{\cos \Psi} \Biggr) \frac{(s+2w)^2 \varepsilon_r^{3/2}}{c^3K(\sqrt{1-k^2})K(k)}f^3, \\
\end{equation}
where $\alpha$ is the attenuation constant, $\Psi = \arccos(\sqrt{\varepsilon_\mathrm{eff}/\varepsilon_\mathrm{r}})$ is the radiation angle\cite{10.1063/5.0005047} (see Fig. \ref{cpw_animation}. $90^\circ - \Psi$ is equivalent to the Mach angle for sonic shock waves), $\varepsilon_\mathrm{eff}$ is the effective dielectric constant of the CPW mode (defined as $\varepsilon_\mathrm{eff} \equiv \sqrt{(c/v_\mathrm{p})}$, where $v_\mathrm{p}$ is the phase velocity), $\varepsilon_\mathrm{r}$ is the relative dielectric constant of the substrate, $s$ is the width of the center strip of the CPW, $w$ is the width of the CPW slots, $c$ is the velocity of light in vacuum, $k=s/(s+2w)$, $f$ is the frequency, and $K$ is the complete elliptical integral of the first kind. It should be noted that the simulated data has an offset from zero loss at zero distance, most likely due to port discontinuities. 

The current distribution around the CPW is shown in Fig.~\ref{cpw_animation} (\href{https://youtu.be/iB3sx6GNrms}{\faYoutubePlay \ \textsf{online video}}). The current propagates away from the CPW, showing no indication of standing waves in the vertical direction. The direction of the shock waves agrees well with $\Psi =  \arccos(\sqrt{\varepsilon_\mathrm{eff}/\varepsilon_\mathrm{r}}) = 42.5^\circ$.

\section{Discussion and Conclusion}

Our results show that the 2.5-dimensional MoM solver \textsf{Sonnet} can be used to accurately simulate the radiation loss from planar superconducting circuits, placed in between a thick substrate and vacuum. The \textsf{Sonnet} simulations of the resonator reproduce the $|S_{31}^2|$ measurement result to within the error margin of the measurement. The \textsf{Sonnet} results are robust against variations in the geometric parameters, though it appears to be that changing the parameters studied here lead to a small systematic \textit{increase} in the $|S_{31}^2|$ peak value. This could be an indication that even better accuracy could be reached by studying the effect of the box size and ground plane size, in combination, in greater detail.

\acknowledgments 
 
AE would like to thank Henry Kool and Andr\'{e} van den Berg at ICT \& Facility Management of TU Delft, for their enthusiastic support in establishing the remote-desktop server used for this research. AE would like to thank Greg Kinnetz at Sonnet for his support in setting up the \textsf{Sonnet}  stratification. AE was supported by the Netherlands Organization for Scientific Research NWO (Vidi grant n$^\circ$ 639.042.423). JJAB was the supported by the European Research Counsel ERC (ERC-CoG-2014 - Proposal n$^\circ$ 648135 MOSAIC). 

\bibliography{ms} 

\begin{thebibliography}{10}

\bibitem{2019NatAs.tmp..418E}
{Endo}, A., {Karatsu}, K., {Tamura}, Y., {Oshima}, T., {Taniguchi}, A.,
  {Takekoshi}, T., {Asayama}, S., {Bakx}, T. J.~L.~C., {Bosma}, S., {Bueno},
  J., {Chin}, K.~W., {Fujii}, Y., {Fujita}, K., {Huiting}, R., {Ikarashi}, S.,
  {Ishida}, T., {Ishii}, S., {Kawabe}, R., {Klapwijk}, T.~M., {Kohno}, K.,
  {Kouchi}, A., {Llombart}, N., {Maekawa}, J., {Murugesan}, V., {Nakatsubo},
  S., {Naruse}, M., {Ohtawara}, K., {Pascual Laguna}, A., {Suzuki}, J.,
  {Suzuki}, K., {Thoen}, D.~J., {Tsukagoshi}, T., {Ueda}, T., {de Visser},
  P.~J., {van der Werf}, P.~P., {Yates}, S. J.~C., {Yoshimura}, Y.,
  {Yurduseven}, O., and {Baselmans}, J. J.~A., ``{First light demonstration of
  the integrated superconducting spectrometer},'' {\em Nature Astronomy}~{\bf
  3},  989--996 (Aug. 2019).

\bibitem{Rantwijk2016}
{van Rantwijk}, J., {Grim}, M., {van Loon}, D., {Yates}, S., {Baryshev}, A.,
  and {Baselmans}, J., ``Multiplexed readout for 1000-pixel arrays of microwave
  kinetic inductance detectors,'' {\em IEEE Transactions on Microwave Theory
  and Techniques}~{\bf 64}(6),  1876--1883 (2016).

\bibitem{2011ApJS..194...24M}
{Monfardini}, A., {Benoit}, A., {Bideaud}, A., {Swenson}, L., {Cruciani}, A.,
  {Camus}, P., {Hoffmann}, C., {D{\'e}sert}, F.~X., {Doyle}, S., {Ade}, P.,
  {Mauskopf}, P., {Tucker}, C., {Roesch}, M., {Leclercq}, S., {Schuster},
  K.~F., {Endo}, A., {Baryshev}, A., {Baselmans}, J.~J.~A., {Ferrari}, L.,
  {Yates}, S.~J.~C., {Bourrion}, O., {Macias-Perez}, J., {Vescovi}, C.,
  {Calvo}, M., and {Giordano}, C., ``{A Dual-band Millimeter-wave Kinetic
  Inductance Camera for the IRAM 30 m Telescope},'' {\em Astrophysical Journal
  Supplement Series}~{\bf 194},  24 (June 2011).

\bibitem{10.1038/nature13171}
Barends, R., Kelly, J., Megrant, A., Veitia, A., Sank, D., Jeffrey, E., White,
  T.~C., Mutus, J., Fowler, A.~G., Campbell, B., Chen, Y., Chen, Z., Chiaro,
  B., Dunsworth, A., Neill, C., O’Malley, P., Roushan, P., Vainsencher, A.,
  Wenner, J., Korotkov, A.~N., Cleland, A.~N., and Martinis, J.~M.,
  ``{Superconducting quantum circuits at the surface code threshold for fault
  tolerance},'' {\em Nature}~{\bf 508}(7497),  500--503 (2014).

\bibitem{doi:10.1063/1.4919761}
Bruno, A., de~Lange, G., Asaad, S., van~der Enden, K.~L., Langford, N.~K., and
  DiCarlo, L., ``Reducing intrinsic loss in superconducting resonators by
  surface treatment and deep etching of silicon substrates,'' {\em Applied
  Physics Letters}~{\bf 106}(18),  182601 (2015).

\bibitem{PhysRevLett.109.107003}
Driessen, E. F.~C., Coumou, P. C. J.~J., Tromp, R.~R., de~Visser, P.~J., and
  Klapwijk, T.~M., ``Strongly disordered tin and nbtin $s$-wave superconductors
  probed by microwave electrodynamics,'' {\em Phys. Rev. Lett.}~{\bf 109},
  107003 (Sep 2012).

\bibitem{sonnet_user_guide}
``{Sonnet User's Guide | Sonnet }.''
  \url{http://www.sonnetsoftware.com/support/manuals.asp}.

\bibitem{CSTMicrowaveStudio}
``{CST Microwave Studio-3D EM simulation software}.''
  \url{https://www.cst.com/products/cstmws}.

\bibitem{HFSS}
``{High Frequency Structure Simulator (HFSS)}.''
  \url{https://www.ansys.com/products/electronics/ansys-hfss}.

\bibitem{sonnet_how_EM_works}
``{How EM Works | Sonnet }.''
  \url{http://www.sonnetsoftware.com/products/sonnet-suites/how-EM-works.html}.

\bibitem{10.1063/5.0005047}
H\"ahnle, S., Marrewijk, N.~v., Endo, A., Karatsu, K., Thoen, D.~J., Murugesan,
  V., and Baselmans, J. J.~A., ``{Suppression of radiation loss in high kinetic
  inductance superconducting co-planar waveguides},'' {\em Applied Physics
  Letters}~{\bf 116}(18),  182601 (2020).

\bibitem{2012ITMTT..60.1235N}
{Noroozian}, O., {Day}, P.~K., {Eom}, B.~H., {Leduc}, H.~G., and {Zmuidzinas},
  J., ``{Crosstalk Reduction for Superconducting Microwave Resonator Arrays},''
  {\em IEEE Transactions on Microwave Theory Techniques}~{\bf 60},  1235--1243
  (May 2012).

\bibitem{Wheeler:2018cg}
Wheeler, J., Hailey-Dunsheath, S., Shirokoff, E., Barry, P.~S., Bradford,
  C.~M., Chapman, S.~C., Che, G., Doyle, S., Glenn, J., Gordon, S., Hollister,
  M.~I., Kov{\'a}cs, A., LeDuc, H.~G., Mauskopf, P.~D., McGeehan, R., McKenney,
  C., Reck, T.~J., Redford, J., Ross, C., Shiu, C., Tucker, C., Turner, J.,
  Walker, S., and Zmuidzinas, J., ``{SuperSpec, The On-Chip Spectrometer:
  Improved NEP and Antenna Performance},'' {\em J. Low Temp. Phys.}~{\bf 84},
  1--7 (May 2018).

\bibitem{2019JATIS...5c5004E}
{Endo}, A., {Karatsu}, K., {Laguna}, A.~P., {Mirzaei}, B., {Huiting}, R.,
  {Thoen}, D.~J., {Murugesan}, V., {Yates}, S. J.~C., {Bueno}, J., {Marrewijk},
  N.~v., {Bosma}, S., {Yurduseven}, O., {Llombart}, N., {Suzuki}, J., {Naruse},
  M., {de Visser}, P.~J., {van der Werf}, P.~P., {Klapwijk}, T.~M., and
  {Baselmans}, J. J.~A., ``{Wideband on-chip terahertz spectrometer based on a
  superconducting filterbank},'' {\em Journal of Astronomical Telescopes,
  Instruments, and Systems}~{\bf 5},  035004 (Jul 2019).

\bibitem{plotting_conventions}
``{Plotting Conventions | Sonnet }.''
  \url{http://www.sonnetsoftware.com/support/sonnet/help_topics/plotting_considerations.htm}.

\bibitem{button1983infrared}
Button, K.~J.,  [{\em Infrared and Millimeter Waves V10: Millimeter Components
  and Techniques, Part II}{\nolinebreak\hspace{0.1em}]}, Elsevier (1983).

\bibitem{81658}
{Frankel}, M.~Y., {Gupta}, S., {Valdmanis}, J.~A., and {Mourou}, G.~A.,
  ``Terahertz attenuation and dispersion characteristics of coplanar
  transmission lines,'' {\em IEEE Transactions on Microwave Theory and
  Techniques}~{\bf 39}(6),  910--916 (1991).

\end{thebibliography}
\bibliographystyle{spiebib} 

\end{document}